\documentclass[fleqn,twoside]{article}
\usepackage{espcrc2}

\usepackage{graphicx}
\usepackage[figuresright]{rotating}

\newcommand{\AmS}{{\protect\the\textfont2
  A\kern-.1667em\lower.5ex\hbox{M}\kern-.125emS}}

\newcommand{\psla}{p\kern-.45em/}
\newcommand{\esla}{E\kern-.45em/}

\newcommand{\beq}{ \begin{eqnarray} }
\newcommand{\eeq}{ \end{eqnarray} }
\newcommand{\beqstar}{ \begin{eqnarray*} }
\newcommand{\eeqstar}{ \end{eqnarray*} }
\newcommand{\gsim}{ \mathop{}_{\textstyle \sim}^{\textstyle >} }
\newcommand{\lsim}{ \mathop{}_{\textstyle \sim}^{\textstyle <} }

\title{Probing physics beyond the standard model from lepton sector}

\author{J. Hisano\address[ICRR]{
Institute of Cosmic Ray Research, University of Tokyo, 
 Kashiwa 277-8582, Japan }
\address[KEK]{Theory Group, KEK, Tsukuba 305-0801, Japan}
}       
\begin{document}

\begin{abstract}

In this review we discuss physics of the lepton sector, the anomalous
dipole moment of muon, the charged lepton-flavor violation, and the
electric dipole moments of charged leptons, from viewpoints of the
minimal supersymmetric standard model and the extensions.

\vspace{1pc}
\end{abstract}

\maketitle

\section{Introduction}

The Standard Model (SM) is the most successful model to explain
physics below the weak scale. The recent results for $\sin 2\beta$ or
$\sin 2\phi_1$ by Belle \cite{belle} and Babar \cite{babar} are
converging, and they are consistent with the Kobayashi-Maskawa
mechanism \cite{KM}. Also, the precision measurements of the
electroweak parameters suggest the light SM Higgs boson such as $m_h
\lsim 196$GeV (95\%CL) \cite{lephiggs}. Now we are waiting for signature
of physics beyond the SM.
 
We have the clue for the physics beyond the SM in the neutrino
oscillation experiment results. The atmospheric neutrino result by the
superKamiokande experiment is established \cite{skatm}.  The combined
result of the superKamiokande
\cite{sksol} and SNO
\cite{sno} experiments shows a strong evidence for the appearance of
$\nu_\mu$ or $\nu_\tau$ in the solar neutrino. At present, the large-angle
MSW solution is the most favored, which will be checked by the Kamland
experiment soon \cite{kamland}.  These neutrino oscillation results
can be explained by introduction of small neutrino masses.

Introduction of small neutrino masses gives us a new scenery for
physics beyond the SM. While the promising candidate for origin of the
small neutrino masses is the see-saw mechanism \cite{seesaw} by
introduction of the right-handed neutrinos, the atmospheric neutrino
result implies that the right-handed neutrino mass should be smaller
than $\sim 10^{-15}$GeV, which is much smaller than the Planck
scale. This is a good motivation for introduction of Supersymmetry
(SUSY). A quadratically-divergent contribution to the Higgs boson mass
proportional to the right-handed neutrino masses should be canceled
even if the correction proportional to the Planck mass is vanishing by
some mysterious physics.

Nowadays, the minimal supersymmetric standard model (MSSM) is one the
most promising extension of standard model (SM), and many experiments
are searching for the possible evidence of the low-energy
supersymmetry. In this review we discuss the possibilities in physics
of charged lepton, paying attention to the dipole-moment operators;
\begin{eqnarray}
{\cal L}
&=&
e \frac{
m_{l_j}}{2} \bar{l_i} 
\sigma_{\mu\nu} F^{\mu \nu}
(L_{ij} P_L+ R_{ij} P_R) l_j
\label{dipole}
\end{eqnarray}
where $P_{R/L} = (1 \pm \gamma_5)/2$, and $i,j$ are for the
generation.  These operators are sensitive to physics beyond the
SM. The real diagonal parts of $L_{ij}$ and $R_{ij}$ contributes to
the anomalous magnetic moments of charged leptons, $a_{l_i}(\equiv
(g_{l_i}-2)/2)) = m_{l_i}^2 (R_{ii} + L_{ii})$. They are sensitive to
structure of the Higgs sector, since the operator (\ref{dipole}) is
violating the lepton chiral symmetry and the SU(2)$_L\times$U(1)$_Y$
symmetry. In fact, the MSSM may predict the larger correction to it
than the electroweak correction since the MSSM has two doublet Higgs
bosons.

If non-vanishing $L_{ij}$ or  $R_{ij}$ ($i\ne j$) exits, the
charged lepton-flavor violating (LFV) processes, such as
$\mu\rightarrow e \gamma$, are predicted;
$Br(\mu\rightarrow e \gamma) 
\propto(|R_{\mu e}|^2+ |L_{\mu e}|^2)$.
Now we know from the neutrino oscillation results that the
lepton-flavor symmetry is not exact in nature, and the problem is how
large is the charged LFV. The small neutrino masses themselves,
expected from the neutrino oscillation results, cannot give any prediction
for the charged LFV processes accessible in near future. In the MSSM,
the charged LFV is supplied by the SUSY breaking masses of sleptons,
and the magnitude depends on the origin of the SUSY breaking and
interaction beyond the MSSM, such as in see-saw mechanism or the
supersymmetric grand unified models (SUSY GUTs).

When diagonal parts of $L_{ij}$ and/or $R_{ij}$ have imaginary part,
CP is violating and the electric dipole moments (EDM) are predicted;
$d_{l_i} = e m_{l_i} {\rm Im} (R_{ii}-L_{ii})$. The EDMs are also
supplied by the SUSY breaking slepton masses in the MSSM.

We organize this review as follows. In the next section we summarize
the current status of the muon $(g-2)$. In section 2 we discuss
dependence of the charged LFV processes on the SUSY breaking models,
and show the branching ratios of the charged LFV processes in the
supersymmetric see-saw model, using the neutrino oscillation data.
Section 3 is for the EDMs of charged leptons. Section 4 is summary.

\section{Muon anomalous magnetic moment}

The latest result for the anomalous magnetic moment of muon
(BNL'98\&'99+CERN'77 \cite{g-2bnl}) is $a_{\mu}^{\rm exp} = (116\ 592\
023 \pm 151) \times 10^{-11}$, while the SM prediction is
$a_{\mu}^{\rm SM} = (116\ 592\ 768 \pm 65) \times 10^{-11}.$ The
contents of the SM contribution are listed in Table~1. The deviation
of the measurement from the SM prediction is $ a_{\mu}^{\rm NP}
(\equiv a_\mu^{\rm exp}-a_\mu^{\rm SM}) = 255\pm 164\times 10^{-11}$,
and it is $1.6\sigma$ away. At present the significance of the
deviation is small. The experimental error is expected to be improved
by a factor 2 in BNL'00 data, and the ultimate goal may be $\sim
40\times 10^{-11}$.
\begin{table}[htb]
\caption{The SM contribution to the muon $g-2$.}
\begin{tabular}{|l|c|c|}\hline
   &   $(\times10^{11})$ & \\
\hline
$a^{\rm QED}_\mu$ & 116 584 705.7(2.9) &\cite{qedg-2} \\
$a^{\rm Had}_\mu$(VP1) & 6 924(62) &\cite{Davier:1998si} \\
$a^{\rm Had}_\mu$(VP2) & -100(6) & \cite{ADH}\\
$a^{\rm Had}_\mu$(LbyL) & 86(19)&\cite{Hayakawa:2001bb} \cite{Bijnens:2001cq}\\
$a^{\rm EW}_\mu$(1 loop) & 195 &\cite{ewg-2} \\
$a^{\rm EW}_\mu$(2 loop) & $-43$(4) &\cite{ewg-22} \\
\hline
$a^{\rm SM}_\mu$ & 116 591 768(65)& \\
\hline
\end{tabular}
\end{table}

Before going to the SUSY contribution to the muon $g-2$, we review the
error in the SM prediction.  The largest ambiguities in the SM
prediction come from the leading hadronic vacuum polarization
contribution $a^{\rm Had}_\mu$(VP1) and the hadronic light-by-light (LbyL)
scattering contribution $a^{\rm Had}_\mu$(LbyL).  The $a^{\rm
Had}_\mu$(VP1) in Table~1 is derived by M. Davier and A. Hocker
\cite{Davier:1998si} from the $e^+e^-$ hadronic
cross section and the hadronic $\tau$ decay data, including
perturbative QCD calculation in the high $q^2$ part.\footnote{ 
See also Refs.~\cite{hadrong-2} for  discussion of the
estimation of the leading hadronic vacuum polarization contribution.  
} 
This estimation will be further improved by high quality data for the
$e^+e^-$ hadronic cross section by CMD2 in Novosibirsk \cite{cdm2},
KLOE in Fascati \cite{kloe}, and BES in Beijing \cite{bes}.  The Babar
may contribute to it by measurement via the initial state radiation of
hard photon \cite{babarg-2}. The CLEO and LEP data for the tau
decay are also important.

On the other hand, the estimate of the LbyL scattering contribution
relies on the model calculation.  The LbyL contribution comes
from three type diagrams in Fig.~1, and the value in Table~1 is the
average value for the latest results of Refs.~\cite{Hayakawa:2001bb} and
\cite{Bijnens:2001cq};
\begin{eqnarray}
a^{\rm Had}_\mu\mbox{(LbyL)} &=& (89.6\pm 15.4)\times 10^{-11}~
\cite{Hayakawa:2001bb}, 
\label{HK}\\
a^{\rm Had}_\mu\mbox{(LbyL)} &=& (83\pm 32)\times 10^{-11}~
\cite{Bijnens:2001cq}.
\label{bijnes}
\end{eqnarray}
The dominant contribution in $a^{\rm Had}_\mu\mbox{(LbyL)}$ comes from
the pion-pole diagram. This diagram was reevaluated by several groups
\cite{lbyl}\cite{Hayakawa:2001bb}\cite{Bijnens:2001cq}, and the sign
problem has been fixed now. However, still they rely on the model
calculation since the diagram is divergent. They are based on the
chiral perturbation or the ENJL model.  The vector-meson dominance is
assumed and the phenomenological parametrization of the pion form
factor $\pi\gamma^*\gamma^*$ is introduced in order to regularize the
divergence.
\begin{figure}[htb]
\begin{center}
\includegraphics[width=16pc]{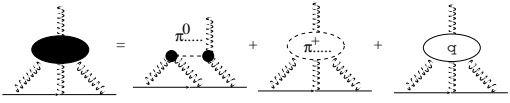}
\end{center}
\caption{The light-by-light scattering contributions to the muon $g-2$.}
\label{fig_ll}
\end{figure}

In Ref.~\cite{Ramsey-Musolf:2002cy} the pion-pole contribution 
is evaluated in a model-dependent way, based on the chiral perturbation 
theory. The result is following;
\begin{eqnarray}
&a^{\rm Had}_\mu\mbox{(LbyL)}|_{\pi^0 {\rm pole}} =
\frac3{16\pi^2}
\left(\frac{\alpha}{\pi}\right)^3
\left(\frac{m_\mu}{F_\pi}\right)^2&
\nonumber\\
&
\times\left\{
\log^2\frac{\Lambda}{m_\mu}
+(\frac16 \chi-0.17) \log\frac{\Lambda}{m_\mu}
+\tilde{C}
\right\},&
\nonumber
\end{eqnarray}
where $\Lambda$ is the ultraviolet cutoff ($\Lambda\sim 4\pi F_\pi$).
The largest term proportional to $\log^2{\Lambda}/{m_\mu}$ is fixed by
the gauge invariance and chiral anomaly.  $\chi$ is
a counter term to regularize the two-loop diagram. While it can be
determined by the leptonic decay of the psuedescalar mesons, the
sensitivity is low at present.  Furthermore, $\tilde{C}$, which is
a piece not enhanced by $\log$, cannot be evaluated without
explicit models. The uncertainty due to $\tilde{C}$  is 
$\delta a_\mu = 31 \times 10^{-11} \tilde{C}$. 

While the model-dependent calculations (\ref{HK}) and (\ref{bijnes})
seem to be converged, we do not have a strategy to derive the
pion-pole contribution precisely enough in a model-dependent way.
Also, we have a subtle problem in the light-by-light contribution,
whether the inclusion of the quark loop is double-counting or not.
Thus, the calculation of the light-by-light contribution on base of
QCD is strongly desired.

\begin{figure}[htb]
\begin{center}
\includegraphics[width=16pc]{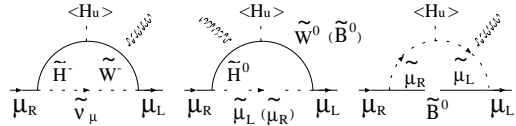}
\end{center}
\caption{The contributions in the MSSM to the muon $g-2$,
which are enhanced by $\tan\beta$.}
\label{fig2}
\end{figure}

If the hadronic contribution is well-controlled, the muon $g-2$ is
so sensitive to physics beyond the SM \cite{Czarnecki:2001pv}, as
mentioned in Introduction.  Before closing this section, we discuss it
from a viewpoint of the MSSM. The nature of two Higgs doublet model
in the MSSM can enhance the contribution, and the contribution
proportional to $\tan\beta$ \cite{SUSYg-2}, which comes from
Fig.~(\ref{fig2}), is given as
\begin{eqnarray}
a_\mu^{\rm SUSY} &\simeq& 
\frac{5\alpha_2+\alpha_Y}{48\pi} \frac{m_\mu^2}{m_{\rm S}^2} 
{\tan\beta}
\nonumber\\
&\simeq&
1.3\times 10^{-9} \left(\frac{100{\rm GeV}}{m_{\rm SUSY}}\right)^2 \tan\beta.
\nonumber
\end{eqnarray}
Here, all relevant SUSY breaking parameters are assumed to be common to
$m_{S}$.  Thus, it may be larger than the electroweak correction in
the SM and the deviation from the SM in the MSSM may reach to
$10^{-8}$.

If the sizable deviation is observed, it will supply a big impact
on  the model building and the phenomenology of the MSSM. After 
``observation'' of $2.6\sigma$ deviation of the muon $g-2$, there were
so many activities for it. Here we summarize them briefly.

First is for the mass spectrum of the SUSY particles in the MSSM.  If the
deviation is observed, relatively lighter slepton and chargino or
larger $\tan\beta$ will be favored
\cite{afterg-21}\cite{afterg-22}. Especially, when the Higgs mass constraint
is included, the SUSY particle spectrum is more constraint, since it
prefers to the large stop mass and/or large $\tan\beta$
\cite{higgslimit}. In the decoupling solution for the FCNC
problem \cite{effsusy}, the first and second generation sfermions are
much heavier than the weak scale. In these models, even if the slepton mixing
between the second and third generations is introduced, the
deviation of the muon $g-2$ can reach to $10^{-9}$ at most
\cite{Cho:2001hx}.

Second is related to sign of the Higgsino mass parameter $\mu$. The muon $g-2$
contribution in the MSSM is proportional to $\mu M_2$ in the broad
parameter space. Here $M_2$ is the SU(2)$_L$ gaugino mass. Now
$b\rightarrow s
\gamma$ favors $\mu A_t>0$, where $A_t$ is the 
$\tilde{t}_L$-$\tilde{t}_R$-Higgs trilinear coupling. If the SUSY
breaking in the MSSM comes from physics at high energy scale, such as
the minimal supergravity model, $A_t\sim M_3$ where $M_3$ is the
SU(3)$_C$ gaugino mass. If sign of the two gaugino masses is required
to have the same sign, the anomaly mediated SUSY breaking model
\cite{anomaly} will be
disfavored \cite{afterg-21}. The consistency of the muon $g-2$ and the
Yukawa unification ($Y_t=Y_b=Y_\tau$) at the GUT scale is also
interesting since the Yukawa unification favors $\mu M_3<0$ \cite{yukawa}.

Third is the $\tan\beta$ enhanced processes.  If the muon $g-2$
deviation is observed, it will give normalizations of the processes induced
by dipole operators. Especially, the LFV processes, such as
$\mu\rightarrow e\gamma$, have a direct relation to it
\cite{HT}. The processes generated by the Yukawa coupling
may be also enhanced. For example, the counting rate of the neutralino 
dark matter would be enhanced \cite{Baltz:2001ts}.

\section{Lepton-flavor violation in the charged-lepton sector}

While the lepton-flavor violation is observed in the neutrino
oscillation experiments, this does not mean sizable LFV processes in
the charged-lepton sector exit. The charged LFV processes induced by
the small neutrino masses, expected from the neutrino oscillation results, are
suppressed by the GIM mechanism, as $Br(\mu\rightarrow e\gamma)
\lsim 10^{-48}({m_{\nu}}/{1{\rm eV}})^4$, even if the neutrino mixing is  maximal.
On the other hand, if the SM is supersymmetrized, the situation is
changed.  The SUSY breaking slepton masses are not necessary aligned
to the lepton masses, and it may lead to sizable lepton-flavor
violating.

\begin{figure}[htb]
\begin{center}
\includegraphics[width=16pc]{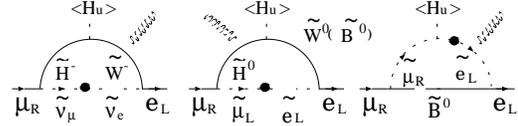}
\end{center}
\caption{The contributions in the MSSM to the $\mu\rightarrow e \gamma$,
which are enhanced by $\tan\beta$. Here, it is assumed that the
left-handed sleptons have the LFV masses.}
\label{fig3}
\end{figure}

Let us asuume that $(m_{\tilde{L}}^2)_{12}$ in the left-handed slepton
mass matrix is non-vanishing. In this case, $\mu\rightarrow e \gamma$
is generated by diagrams in Fig.~(\ref{fig3}), and the approximate
formula is given as $Br(\mu\rightarrow e \gamma)
\simeq
3\times10^{-5}({a_\mu^{\rm SUSY}}/{10^{-9}})^
2({(m_{\tilde{L}}^2)_{12}}/{m_{S}^2})^2$ \cite{HT}.
The diagrams in Fig~(\ref{fig3}) are so similar to the diagrams in 
Fig.~(\ref{fig2}) contributing to the muon $g-2$, and the muon $g-2$
gives the normalization of the branching ratio of $\mu\rightarrow e \gamma$.

In Table~2 we summarize the current experimental bounds to the charged
LFV processes, the sensitivities in the present activities, and
prospects in the future experiments such as the PRISM project
\cite{PRISM} and the front ends of neutrino factories under
consideration at CERN \cite{nf}. The charged LFV processes are
radiative-induced in the MSSM as far as the R party is not broken.
Thus, the branching ratio of $\mu \rightarrow 3e$ and the $\mu$-$e$ 
conversion rate in nuclei are approximately given  as
${Br(\mu \rightarrow 3e)}/{Br(\mu \rightarrow e \gamma)}
\simeq  7 \times 10^{-3}$ and 
${R(\mu^- {\rm Ti(Al)}\rightarrow e^- {\rm Ti(Al)})}/
{Br(\mu\rightarrow e \gamma)}
\simeq 
5(3) \times 10^{-3}$. (See the detailed calculation of the $\mu$-$e$
conversion rate in nuclei is given in Refs.~\cite{Kitano:2002mt}.)
From these simple formulas, the naive current bound on
$(m_{\tilde{L}}^2)_{12}/{m_{S}^2}$ is $\lsim {6
\times10^{-4} } ({\delta a_\mu^{\rm SUSY}}/{10^{-9}})^{-1}$, and 
PSI and MECO/BNL (PRISM and NuFACT) may reach to $\sim
{10^{-5}}({10^{-6}})$.  These experiments are stringent tests of the
lepton-flavor symmetry in the MSSM.
\begin{table*}[htb]
\caption{Current experimental bounds to the charged
LFV processes, the sensitivities in the present activities, and
prospects in the future experiments.}
\begin{tabular}{||l|c|c|c||}\hline
&Current bound&Present Activities & Future \\
\hline
$\tau\rightarrow \mu \gamma$&
$1.0\times 10^{-6}$ \cite{belltau}& 
$\sim 10^{-7}$ (Belle/KEK) \cite{ohshima}&
$10^{-(8-9)}$ \cite{ohshima}
\\
\hline
$\mu\rightarrow e \gamma$&
$1.2\times 10^{-11}$ \cite{Brooks:1999pu}& 
$2\times 10^{-14}$ (PSI) \cite{PSI} &
$10^{-15}$ \cite{nf}
\\
\hline
$\mu\rightarrow 3 e$&
$1.0\times 10^{-12}$  \cite{PDG} & 
&
$10^{-(15-16)}$ \cite{nf}
\\
\hline
$\mu^- N\rightarrow e^- N$&
$6.1\times 10^{-13}$ \cite{mueconv}& 
\begin{tabular}{c}
$10^{-14}$ for Ti (SINDRUM II) \cite{sindrum2}\\
$5 \times 10^{-17}$ for Al (MECO/BNL) \cite{meco}
\end{tabular}
 &
$10^{-18}$ \cite{nf} \cite{PRISM}\\
\hline
\end{tabular}
\end{table*}

The charged LFV in the MSSM depends on the origin of the SUSY breaking
term in the MSSM and the interaction of physics beyond the MSSM. The
SUSY breaking model is classified to two types by degeneracy or
non-degeneracy of the sfermion masses. The later may predict the large
LFV rates and sometimes the broad parameter region has been excluded
already. Here, we will concentrate on the SUSY breaking models where
the degeneracy of the sfermion masses is predicted by assuming the
hidden sector, such as the gravity-~\cite{grmed},
gauge-~\cite{gaugemed}, anomaly-~\cite{anomaly},
gaugino-mediation~\cite{ginom} models.

The magnitude of the charged LFV processes in these models depends on
the scale of the SUSY breaking mediation ($M_M$) and the scale of the
physics with LFV interaction ($M_{LFV}$). The well-motivated
candidates for the physics with LFV interaction are the SUSY GUTs and
the see-saw mechanism. If $M_M\gsim M_{LFV}$, such as in the
gravity-mediation model, the LFV slepton masses are radiatively
generated by the LFV interaction and they depend on $\log
M_{M}/M_{LFV}$. The LFV processes may have observable rates in this
case \cite{HKR}\cite{nonmgut}\cite{PRL57-961}.

\begin{figure}[htb]
\begin{center}
\includegraphics[width=16pc]{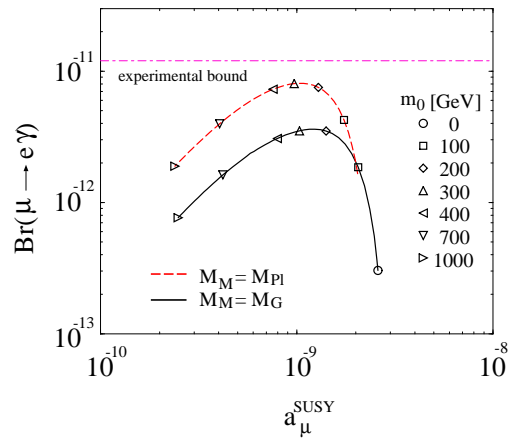}
\end{center}
\caption{Dependence of $Br(\mu\rightarrow e \gamma)$ on 
the universal scalar mass $m_0$ in the supersymmetric seesaw model, assuming
the gravity-mediation model. Here, the SU(2)$_L$ gaugino mass $M_2$
is  $200{\rm GeV}$.}
\label{fig4}
\end{figure}

In the gaugino-mediation model, where $M_M$ is the reduced Planck
scale or the GUT scale, the scalar masses at $M_M$ are vanishing, and
they are generated through the gaugino loops. The LFV slepton masses are
induced at higher order and suppressed. However, the suppression of
the LFV processes is at most a factor 10. In Fig.~(\ref{fig4}) the
dependence of $Br(\mu\rightarrow e
\gamma)$ on the universal scalar mass $m_0$ in the gravity-mediation
model. The two lines are for $M_M$ the GUT scale or the reduced Planck
scale.  Here, the the SU(2)$_L$ gaugino mass $M_2$ is 200GeV, and the
supersymmetric see-saw model is assumed. See Ref.~\cite{HT} for the
other input parameters.  The branching ratio is maximum at $M_2\sim
m_0$. When $m_0$ is zero (the gaugino-mediation limit), the branching
ratio is suppressed, however it is only a factor 10.

When $M_M\lsim M_{LFV}$, such as in the low-energy gauge-mediation
model, the radiative correction is suppressed by a power of
$M_M/M_{LFV}$, and the effect tends to be invisible. The
anomaly-mediation model is exceptional.  While $M_M$ is the
gravitational scale, the SUSY breaking parameters are determined by
only the particle contents and interactions in the MSSM in the
original anomaly-mediation model (the UV insensitivity), and the LFV
slepton masses are suppressed. On the other hand, the model has the
problem of tachyonic sleptons. Then, the LFV slepton masses in this
model depends on how to care of the problem. For example, in the
minimal anomaly-mediation model
\cite{mam}, where the universal scalar mass $m_0$ is added to the
anomaly-mediation contribution, the LFV slepton masses are generated
and proportional to $m_0^2$.

\begin{figure}[htb]
\begin{center}
\includegraphics[width=16pc]{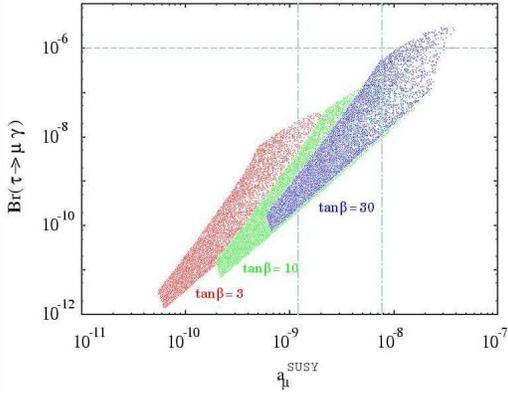}
\end{center}
\caption{$Br(\tau\rightarrow\mu \gamma)$ in the supersymmetric see-saw model,
assuming the gravity-mediation model. Here, we use the atmospheric
neutrino result. The horizontal line is the MSSM contribution to 
the muon $g-2$.}
\label{fig5}
\end{figure}

Next, we will discuss the charged LFV processes in the
supersymmetric see-saw model using the the neutrino oscillation
results. We assume the gravity-mediation model. The atmospheric
neutrino result suggests the large mixing of left-handed stau and
smuon, and it may imply the large branching ratio of
$\tau\rightarrow\mu \gamma$. In Fig.~(\ref{fig5}) we present
$Br(\tau\rightarrow\mu \gamma)$ in this model. Here, we use
$m_{\nu_\tau}^2 =2\times 10^{-3}{\rm eV}^2$ and $U_{ 23}=1/\sqrt{2}$,
and asuume that the large mixing comes from the neutrino Yukawa
coupling and that the Yukawa unification of the tau-neutrino and
top-quark Yukawa couplings is imposed at the reduced Planck
scale. These assumptions make $Br(\tau\rightarrow\mu \gamma)$
enhanced. For the SUSY breaking parameters, we take $m_0<500$GeV, the
U(1)$_Y$ gaugino mass $<500$GeV, and the universal A parameter $A_0$
zero. While a parameter region is excluded, the branching ratio
tends to be smaller than the reach of the KEK Belle. If the large
deviation from the SM prediction in the muon $g-2$, such as $\sim
10^{-9}$, is observed, the branching ratio may be larger than
$10^{-9}$.

\begin{figure}[htb]
\begin{center}
\includegraphics[width=16pc]{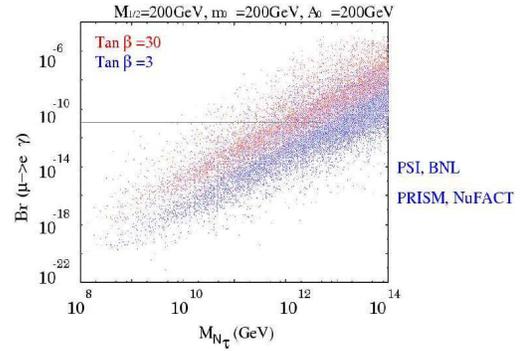}
\end{center}
\caption{
$Br(\mu\rightarrow e \gamma)$ in the supersymmetric see-saw model, using the
atmospheric neutrino result and the large-angle MSW solution.
We assume the gravity-mediation model.}
\label{fig6}
\end{figure}

Now the large-angle MSW solution is the most favored in the solar
neutrino problem, and this may lead to the large branching ratio of
$\mu\rightarrow e \gamma$. In Fig.~(\ref{fig6}) we show
$\mu\rightarrow e \gamma$, using the atmospheric
neutrino result and the large-angle MSW solution; $m_{\nu_\mu}^2
=7.5\times 10^{-5}{\rm eV}^2$ and $U_{ 12}=1/\sqrt{2}$. Here we take
$m_{\nu_e}=0$ and assume the canonical generational structure for the
right-handed neutrino masses. For the SUSY breaking parameters, the
universal gaugino mass $M_{1/2}=200$GeV, $m_0=200$GeV, and
$A_0=200$GeV.  The horizontal line is for the right-handed tau neutrino
mass $M_{N_\tau}$. A broad region has been excluded already, and the
future experiments may cover almost the region above $M_{N_\tau}\gsim
10^{11}$GeV.  If $M_{N_\tau}\lsim$O($10^{11}$)GeV, the Dirac mass for
the tau neutrino is smaller than O(1)GeV, which is much smaller than
the top quark mass.

In this section we discussed the charged LFV. After the SUSY particles
are discovered at LHC or lepton colliders, the LFV slepton decay is
important \cite{ACHF}. The $e^+e^-$ linear collider and muon collider
have sensitivity for the $\tilde{\tau}$-$\tilde{\mu}$ mixing beyond
the current proposed $\tau \rightarrow \mu \gamma$ search.

\section{EDMs}

In this section we discuss the EDMs of charged leptons.  The current
experimental bounds and the sensitivities of the future experiments
are listed in Table~3. While the EDMs are suppressed in the SM as $d_e
(d_\mu) < 10^{-40} (10^{-38})e~{\rm cm}$, they are sensitive to the
MSSM.  The relative phases of the $F$-term SUSY breaking parameters,
the $A$ and $B$ terms and the gaugino masses, contribute to the EDMs.
In this section, we assume for simplicity that the sfermion masses are
flavor-independent and the CP-violating phases of the SUSY breaking
parameters are zero at the SUSY-breaking mediation scale, and consider
the EDMs radiatively-induced in physics beyond the MSSM.

\begin{table*}[htb]
\caption{The current experimental bounds to the electric dipole moments
of charged leptons and the prospects in the future experiments.}
\begin{tabular}{||l|c|c|c||}\hline
&Current bound&Present Activities & Future \\
\hline
$d_e (e$ cm)&
$1.6\times 10^{-27}e$ \cite{edme}&  
&
$10^{-33}$\cite{Lamoreaux:2001hb} \\
\hline
$d_\mu (e$ cm)&
$(3.7\pm3.4)\times 10^{-19}$ \cite{muedm}& 
$2\times 10^{-24}$ (BNL) \cite{muedmbnl}&
$10^{-26}$ \cite{nf} \cite{PRISM} \\
\hline
\end{tabular}
\end{table*}

In the minimal SUSY SU(5) GUT, the predicted EDMs are very small.  The
quark and lepton masses are given by the up-type and down-type quark
Yukawa couplings at the GUT scale.  As the result, the EDMs of
electron and muon are proportional to a Jarskog invariant, $\sim
f_{b}^2 f_{c}^2 f_{t}^4 ~{\rm{Im}} [ V_{11} V_{12}^\star V_{22}
V_{21}^\star ]$, where $V$ is the CKM matrix at the GUT scale. This
situation is similar to the SM. Thus, the EDMs are suppressed so much.

\begin{figure}[htb]
\begin{center}
\includegraphics[width=16pc]{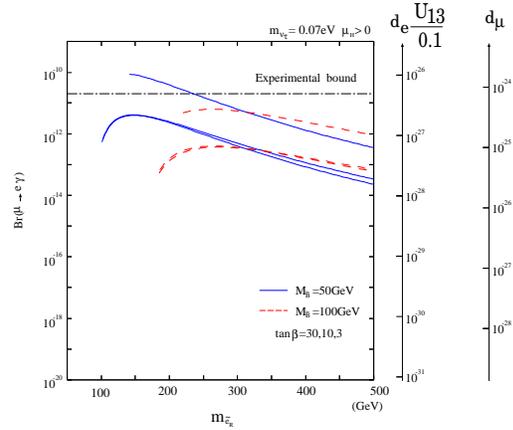}
\end{center}
\caption{$Br(\mu\rightarrow e \gamma)$ and the
EDMs of electron and muon in 
the SUSY SU(5) GUT with the right-handed neutrinos.}
\label{fig7}
\end{figure}

We know that the minimal SUSY SU(5) GUT cannot explain the quark and
lepton masses for the first and second generations, and the extension
is needed. Also, it does not have the right-handed neutrinos. These
extension may change the prediction for the EDM drastically
\cite{nonmgut}. Let us consider that the SUSY SU(5) GUT with the
right-handed neutrinos.  In this case, the EDM of electron (muon) may
be proportional to a Jarskog invariant, $\sim f_{\nu_\tau}^2 f_{t}^2
{\rm{Im}} [ V_{31(2)}V_{33}^\star U_{1(2)3}U_{33}^\star ]$. Here, we
assume for simplicity that the right-handed neutrino masses are
degenerate and $U$ is the MNS matrix at the GUT scale. The relative
phases between $U$ and $V$ contribute to the EDMs.  In
Fig.~(\ref{fig7}) we show the $Br(\mu\rightarrow e \gamma)$ and the
EDMs of electron and muon. We asuume the maximal CP violating
phases. See Ref.~\cite{HNY} for the input parameters in this
figure.  Since the left-handed and right-handed sleptons get the LFV
masses as $(m_{\tilde L}^2)_{ij} \propto U_{i3} U_{j 3}^\star$ and
$(m_{\tilde E}^2)_{ij} \propto V_{3 i} V^{
\star}_{3 j}$, $Br(\mu\rightarrow e \gamma)$ and the EDMs have a
strong correlation.
From this figure it is found that the prediction may be accessible in
the future experiments, and $U_{13}$ is an important parameter for the
electron EDM.

In the supersymmetric see-saw model, if the right-handed neutrino
masses are exactly degenerate, the EDMs of charged leptons are also
suppressed, similar to the minimal SUSY SU(5) GUT. The non-degeneracy
of the right-handed neutrino masses may enhance the EDMs drastically
\cite{edmnu}, and the muon (electron) EDM can reach to
$10^{-26}(10^{-31})e$ cm.

Other CP violating observable in the lepton sector is the T-odd
asymmetry in the polarized muon decay to $3e$. While it comes from
interference between the photon-penguin diagram and the $Z$ penguin
and box diagrams, the photon-penguin diagram tends to be dominant in
the $\mu\rightarrow 3e$ process and the T-odd asymmetry is suppressed.
In the minimal SUSY SU(5) GUT and the supersymmetric see-saw model
the T-odd asymmetry may reach to 10\% if the photon-penguin contribution
is suppressed \cite{Okada}\cite{atnu}.

\section{Summary} 

In this review we discuss physics of the lepton sector from viewpoints
of the minimal supersymmetric standard and the extensions. While the
muon $g-2$ is sensitive to the MSSM, the understanding of the
systematic error in the the SM prediction, especially the
light-by-light contribution, is very serious when the experimental
error is reduced furthermore.  The charged LFV processes depends on
the SUSY breaking models and the LFV interaction beyond the MSSM. The
current neutrino data is encouraging. The interesting future
experiments may give suggestion for the model discrimination. The EDMs
of charged leptons are sensitive to the extension of the SUSY GUTs,
and they may be accessible in the future experiments.

\underline{Acknowledgment} \\
JH thanks  Prof. S.~Ohta for comment on the lattice calculation of the 
light-by-light contribution to the muon $g-2$.

\end{document}